\documentclass[twoside,12pt]{article}
\usepackage{epsf,epsfig,amssymb,amsmath,graphicx,float}
\usepackage{color}

\begin{document}
\renewcommand{\thefootnote}{\fnsymbol{footnote}}

\begin{titlepage}

\begin{center}

\vspace{1cm}

{\Large {\bf Relic Density of Asymmetric Dark Matter in 
Modified Cosmological Scenarios }}

\vspace{1cm}

{\bf Hoernisa Iminniyaz, Burhan Salai, Guoliang Lv}

\vskip 0.15in
{\it
{School of Physics Science and Technology, Xinjiang University, \\
Urumqi 830046, China} \\
}

\abstract{We discuss the relic abundance of asymmetric Dark Matter
  particles in modified cosmological scenarios where the Hubble
  rate is changed with respect to the standard cosmological scenario. The 
  modified Hubble rate leaves its imprint on the relic abundance of asymmetric 
  Dark Matter particles if the asymmetric Dark Matter particles freeze--out in 
  this era. For generality we parameterize the modification of the Hubble rate 
  and then calculate the relic abundance of asymmetric Dark Matter particles 
  and anti--particles. We found the 
  abundances for the Dark Matter particles and anti--particles are enhanced
  in the modified cosmological models. The indirect detection signal is 
  possible for the asymmetric Dark Matter particles due to the increased 
  annihilation rate in the modified cosmological models. Applying Planck data, 
  we find the constraints on the parameters of the modified cosmological 
  models. }
\end{center}
\end{titlepage}
\setcounter{footnote}{0}

\section{Introduction}

Recent Planck team's full sky measurements of Cosmic Microwave Background 
(CMB) temperature anisotropies provide the precise values for the cosmological 
parameters, the amount of cold Dark Matter (DM) relic density is given as
\begin{eqnarray} \label{eq:pdata}
  \Omega_{\rm DM} h^2 = 0.1199 \pm 0.0022\, ,
\end{eqnarray}
where $h = 0.673 \pm 0.098$ is the present Hubble expansion rate in units of 
100 km s$^{-1}$ Mpc$^{-1}$ \cite{Ade:2015xua}.

Although the precise value of the Dark Matter relic density is given by the 
observations, the nature of Dark Matter is still a mysterious  
question for scientists. Experimental searches are ongoing to find the Dark
Matter using direct detection or indirect detection. Parallel to the
experiments, there are also many theoretical works which attempts to disclose 
the character of Dark Matter. 
One of the general belief is that the Weakly Interacting Massive Particles 
(WIMPs) are the good candidates for Dark Matter. Majorana particle neutralino 
is one of the favorite WIMP which is appeared in supersymmetry and its 
particle and anti--particle are the same.  
Because we have no evidence that shows that the Dark Matter should be Majorana
particle, we still have other possibilities that the Dark Matter can
be Dirac particles for which its particle and anti--particle are
different \cite{adm-models,frandsen}. The relic density of asymmetric Dark 
Matter is calculated in the standard cosmological scenario 
\cite{GSV,Iminniyaz:2011yp}. In the standard model of cosmology, it is assumed
that Dark Matter particles were in thermal and chemical equilibirum in the 
radiation--dominated epoch after the period of last entropy production. 
Dark Matter particles decouple when they become nonrelativistic while the 
universe cools down. For asymmetric Dark Matter, the 
asymmetry is created well before the decoupling of the asymmetric Dark Matter 
particles according to the assumption, therefore there are more particles 
than the anti--particles in the beginning. Dark Matter anti--particles are 
annihilated away with the particles and the rest are the Dark Matter particles 
in the present universe. The direct ditection can be used to find the 
asymmetric Dark Matter particles in principle.  

On the other hand, there is no observational evidence before Big Bang 
Nucleosynthesis (BBN). The alternative cosmological models like scalar--tensor 
gravity \cite{Catena}, quintessence models with a kination phase 
\cite{Salati:2002md,Rosati:2003cu,Profumo:2003hq,Pallis:2005hm}, brane world
cosmology 
\cite{Randall:1999vf,Stoica:2000ws,Abou El Dahab:2006wb,Okada:2004nc} or 
the late 
decaying of inflaton model \cite{Arbey:2011gu,Arbey:2008kv,DEramo:2017gpl} and 
anistropic expansion \cite{Kamionkowski,Barrow:1982ei} 
predict that the expansion rate may have been different from the one expected 
by the standard Friedmann--Robertson--Walker model before BBN. If this occurs 
around the time when the asymmetric Dark Matter particles decouples from the 
thermal equilibrium, the relic abundance of asymmetric Dark Matter 
particles are affected by the modified expansion rate of the universe.
The papers
\cite{Iminniyaz:2013cla,Gelmini:2013awa,Wang:2015gua,Meehan:2014zsa,
Iminniyaz:2015wva,Iminniyaz:2016iom}
already discussed the effects of the modified expansion rate on the relic 
density of asymmetric Dark Matter particles. In those papers, 
the authors concluded that the enhanced 
expansion rate leads to the earlier asymmetric Dark Matter 
particles freeze--out and enhanced relic density.  
In the modified cosmological models, the difference of the
abundance between the particle and anti--particle is not large for reasonable
annihilation cross sections. In order to match with the observed range of the
Dark Matter relic density, the annihilation cross section should be large. 
Then the annihilation rate is increased and it leads to the possible
indirect detection signal for asymmetric Dark Matter as well
\cite{Gelmini:2013awa,Meehan:2014zsa,Iminniyaz:2016iom}. 
In this work, we presented the 
generalized parametrization of the modification of the 
expansion rate of the early universe and derived the general equation to
compute the relic density of asymmetric Dark Matter particles. Then we used 
the Planck data to find the constraints on the parameter space of the 
modified cosmological models. 

The next section is devoted to the analysis of the relic density of
asymmetric Dark Matter particles in modified cosmological scenarios both in
numerical and analytical ways. In section 3, using the Planck data, we find
the constraints on the parameter space for the modified cosmological models. 
There is conclusion and summary in the final section.

\section{Relic Density of Asymmetric Dark Matter in Modified Cosmological
  Scenarios}

In this section, we introduce the parameterization of modification of 
the expansion rate of the universe and then derive the generalized equation 
to calculate the relic density for asymmetric Dark Matter. To describe the 
effects of the revised Hubble expansion rate in the early universe, we add 
the new dark density to the Friedmann equation as    
\begin{equation}\label{H_m}
      H^2 = \frac{8\pi G}{3}\, ( \rho_{\rm rad} + \rho_D  ),
\end{equation}
where  
\begin{equation}
      \rho_{\rm rad} = g_* (T) \frac{\pi^2}{30} T^4
\end{equation}
is the radiation energy density with $g_*$ being the effective number
of relativistic degrees of freedom and 
\begin{equation}
      \rho_D = \rho_D(T_0) \left(\frac{T}{T_0} \right)^{n_D},
\end{equation}
here $\rho_D$ parameterizes the modification of the
expansion rate. $n_D$ is the constant which parametrizes the 
behavior of the energy density. We introduce 
\begin{equation}\label{eta}
      \eta \equiv \frac{\rho_D(T_0)}{\rho_{\rm rad}(T_0)},
\end{equation}
where $T_0$ is the reference temperature which approaches to the 
freeze--out temperature, then
\begin{eqnarray}
      H^2  = 
      \frac{8\pi G}{3} \rho_{\rm rad}
      \left[ 1 + \eta \, \frac{g_*(T_0)}{g_*(T)} 
      \left( \frac{T}{T_0} \right)^{n_D-4} 
      \right] \simeq   \frac{8\pi G}{3} \rho_{\rm rad}
       \left[ 1 + \eta \left( \frac{T}{T_0} \right)^{n_D -4} \right].
\end{eqnarray}
The approximation holds in a range of temperatures where $g_*$
does not change sizably with respect to its value at $T_0$.
At BBN time, the radiation energy density remains dominant, and the new term in 
the Hubble rate should vanish in order not to conflict with the BBN prediction.
When $n_D = 6$, the new dark density corresponds to the 
quintessence model with kination phase; and $n_D = 8$ refers to the brane 
world cosmology or the late decaying of inflaton field. 

The number densities of the asymmetric Dark Matter particles are 
resolved by the following Boltzmann equations,
\begin{eqnarray} \label{eq:boltzmann_n}
\frac{{\rm d}n_{\chi}}{{\rm d}t} + 3 H n_{\chi} &=&  - \langle \sigma v\rangle
  (n_{\chi} n_{\bar\chi} - n_{\chi,{\rm eq}} n_{\bar\chi,{\rm eq}})\,;
  \nonumber \\
\frac{{\rm d}n_{\bar\chi}}{{\rm d}t} + 3 H n_{\bar\chi} &=&
   - \langle \sigma v\rangle (n_{\chi} n_{\bar\chi} - n_{\chi,{\rm
       eq}} n_{\bar\chi,{\rm eq}})\,.
\end{eqnarray}
Here only $\chi \bar \chi$ pairs can annihilate into
Standard Model (SM) particles according to the assumption.
$n_{\chi}$, $n_{\bar\chi}$ are the number 
densities of particle and anti--particle. $\langle \sigma v \rangle$
represents the 
product of thermal average of the annihilation cross section and the relative 
velocity of the Dark Matter particle and anti--particle.
The equilibrium number densities of $\chi$ and $\bar{\chi}$ are 
(in equilibrium, the chemical potential of the particles 
$\mu_{\bar\chi} = -\mu_\chi$.)
\begin{equation} \label{n_eq}
  n_{\chi,{\rm eq}} = g_\chi ~{\left( \frac{m T}{2 \pi} \right)}^{3/2}
  {\rm e}^{(-m + \mu_\chi)/T}\,,\,\,\,\,\,\,\,
  n_{\bar\chi,{\rm eq}} =  g_\chi ~{\left( \frac{m T}{2 \pi}
    \right)}^{3/2} {\rm e}^{(-m - \mu_{\chi})/T}\,,
\end{equation}
where $g_{\chi}$ is the number of intrinsic degrees of freedom of the
particles.
New quantities $Y_\chi =n_\chi/s$, $Y_{\bar\chi} = n_{\bar\chi}/s$ and 
$x = m/T$ are introduced in order to solve Eqs.(\ref{eq:boltzmann_n}), where 
\begin{equation}\label{eq:s}
      s= \frac{2 \pi^2 g_{*s}}{45} T^3 
\end{equation}
is the entropy density with the effective number of entropic degrees of 
freedom $g_{*s}$. Assuming that the universe expands adiabatically and using 
the entropy conservation $s R^3$ where $R$ is the scale factor of the universe, 
we obtain 
\begin{equation}\label{eq:sd}
      \frac{ds}{dt} + 3Hs =0\,.
\end{equation}
Inserting Eq.(\ref{eq:s}) into Eq.(\ref{eq:sd}) and using $x = m/T$, we have
\begin{equation}\label{eq:xt}
      \frac{dx}{dt}= \frac{Hx}{1-\frac{x}{3g_{*s}} \frac{dg_{*s}}{dx}}\,,
\end{equation}
assuming $g_*\simeq g_{*s}$ and $dg_{*s}/dx\simeq 0$, we have
\begin{equation}\label{eq:xth}
      \frac{dx}{dt}= Hx\,.
\end{equation}
Then using Eq.(\ref{eq:xth}), with the modified expansion rate, the 
Boltzmann equations (\ref{eq:boltzmann_n}) are rewritten as
\begin{equation} \label{eq:boltzmann_Y}
\frac{d Y_{\chi}}{dx} =
      - \frac{\kappa \langle \sigma v \rangle}{x^2}~
     \left[ 1 + \eta \left( \frac{x_0}{x} \right)^{n_D -4} \right]^{-\frac{1}{2}}
     (Y_{\chi}~ Y_{\bar\chi} - Y_{\chi, {\rm eq}}~Y_{\bar\chi, {\rm eq}}   )\,;
\end{equation}
\begin{equation} \label{eq:boltzmann_Ybar}
\frac{d Y_{\bar{\chi}}}{dx}
    = - \frac{\kappa \langle \sigma v \rangle}{x^2}~
    \left[ 1 + \eta \left( \frac{x_0}{x} \right)^{n_D -4} \right]^{-\frac{1}{2}}
    (Y_{\chi}~Y_{\bar\chi} - Y_{\chi, {\rm eq}}~Y_{\bar\chi, {\rm eq}} )\,,
\end{equation}
where
$\kappa = 1.32\,m M_{\rm Pl}\, \sqrt{g_*}\,$, $M_{\rm Pl}$ is the
reduced Planck mass.
From (\ref{eq:boltzmann_Y}), (\ref{eq:boltzmann_Ybar}), we obtain
\begin{equation} \label{eq:YYbar}
\frac{d Y_{\chi}}{dx} - \frac{d Y_{\bar{\chi}}}{dx} = 0\,.
\end{equation}
This hints that the difference of the abundance of the particles and 
anti--particles are conserved,
\begin{equation}  \label{eq:c}
 Y_{\chi} - Y_{\bar\chi} = \Lambda\,,
\end{equation}
where $\Lambda$ is constant. Then 
Eqs.(\ref{eq:boltzmann_Y}),(\ref{eq:boltzmann_Ybar}) become
\begin{equation} \label{eq:Yc}
\frac{d Y_{\chi}}{dx} =
     - \frac{\kappa \langle \sigma v \rangle}{x^2}~
     \left[ 1 + \eta \left( \frac{x_0}{x} \right)^{n_D -4} \right]^{-\frac{1}{2}}
                 (Y_{\chi}^2 - \Lambda Y_{\chi} - P     )\, ;
\end{equation}
\begin{equation} \label{eq:Ycbar}
\frac{d Y_{\bar{\chi}}}{dx} =
     - \frac{\kappa \langle \sigma v \rangle}{x^2}~
     \left[ 1 + \eta \left( \frac{x_0}{x} \right)^{n_D -4} \right]^{-\frac{1}{2}}   
     (Y_{\bar\chi}^2 + \Lambda Y_{\bar\chi}  - P)\,,
\end{equation}
where  
$P = Y_{\chi, {\rm eq}} Y_{\bar\chi,{\rm eq}} = (0.145g_{\chi}/g_*)^2\, x^3 e^{-2x}$.
The annihilation cross section of WIMP is expanded in the
relative velocity $v$ between the annihilating WIMPs 
in most cases. The thermal average of $\sigma v$ is
\begin{equation} \label{eq:cross}
   \langle \sigma v \rangle = a + 6\,b x^{-1} + {\cal O}(x^{-2})\, ,
\end{equation}
where $a$, $b$ represent the $s$--wave contribution for the 
limit $v\to 0$ and the
$p$--wave contribution for the suppressed $s$--wave annihilation.
\subsection{Numerical solutions}
In the standard frame of the Dark Matter particle
evolution, it 
is assumed the particles $\chi$ and $\bar{\chi}$ were in thermal equilibrium at
high temperature, when the temperature drops below 
the mass ( $T < m$ ) for $m > |\mu_\chi|$, the equilibrium number 
densities for particles and anti--particles are decreased exponentially 
\cite{Iminniyaz:2011yp,standard-cos}. The 
decline of the interaction rates
$\Gamma $ for particle and $\bar{\Gamma}$ for anti--particle
below the Hubble expansion rate $H$ leads to the decoupling of the particles 
and anti--particles from the equilibrium state where the temperature at the 
decoupling is called the freeze--out temperature. After the decoupling, 
the number densities of the particles and anti--particles in a 
co--moving space are nearly constant. 

\begin{figure}[h]
  \begin{center}
    \hspace*{-0.5cm} \includegraphics*[width=8cm]{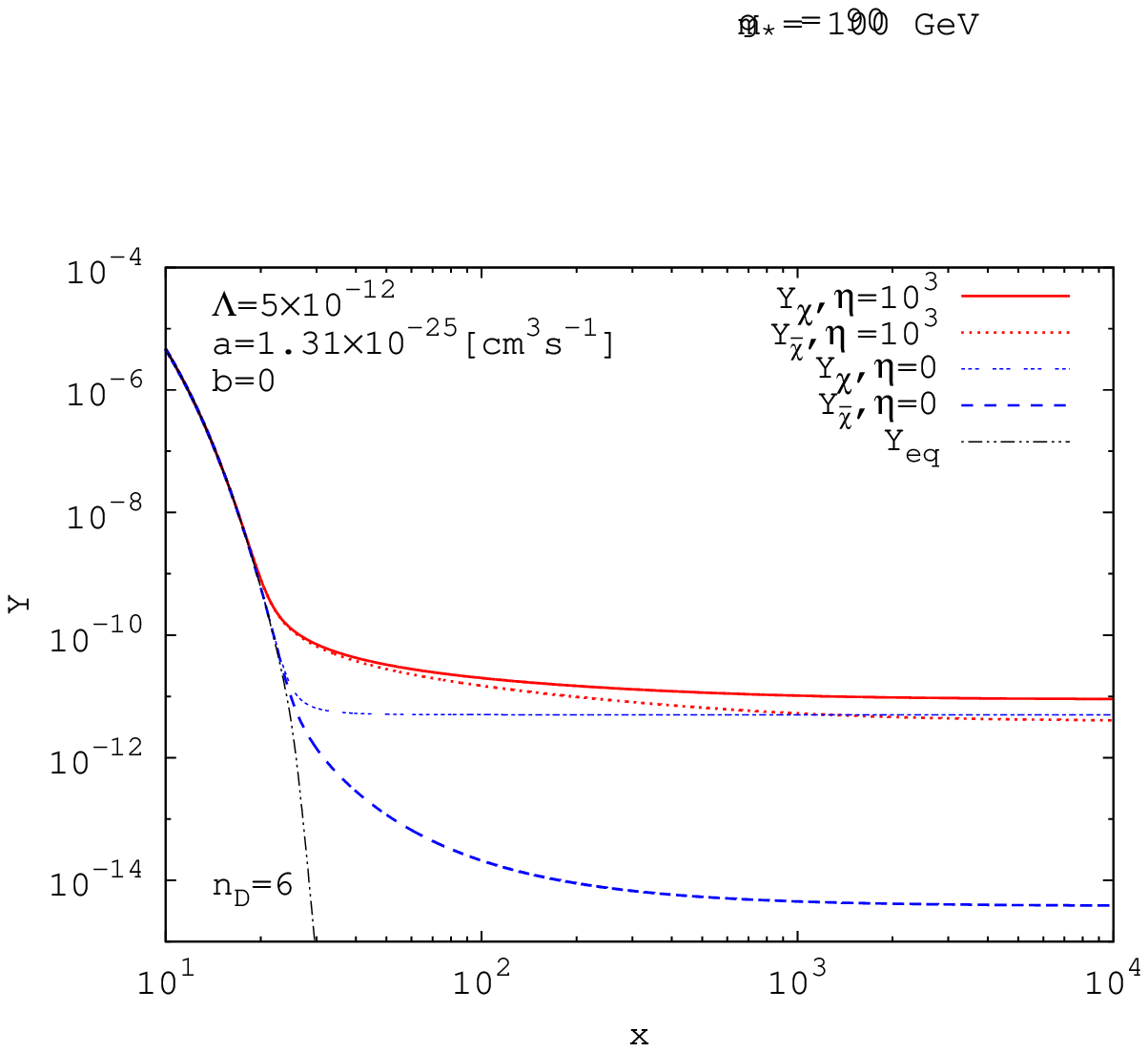}
    \put(-115,-12){(a)}
    \hspace*{-0.5cm} \includegraphics*[width=8cm]{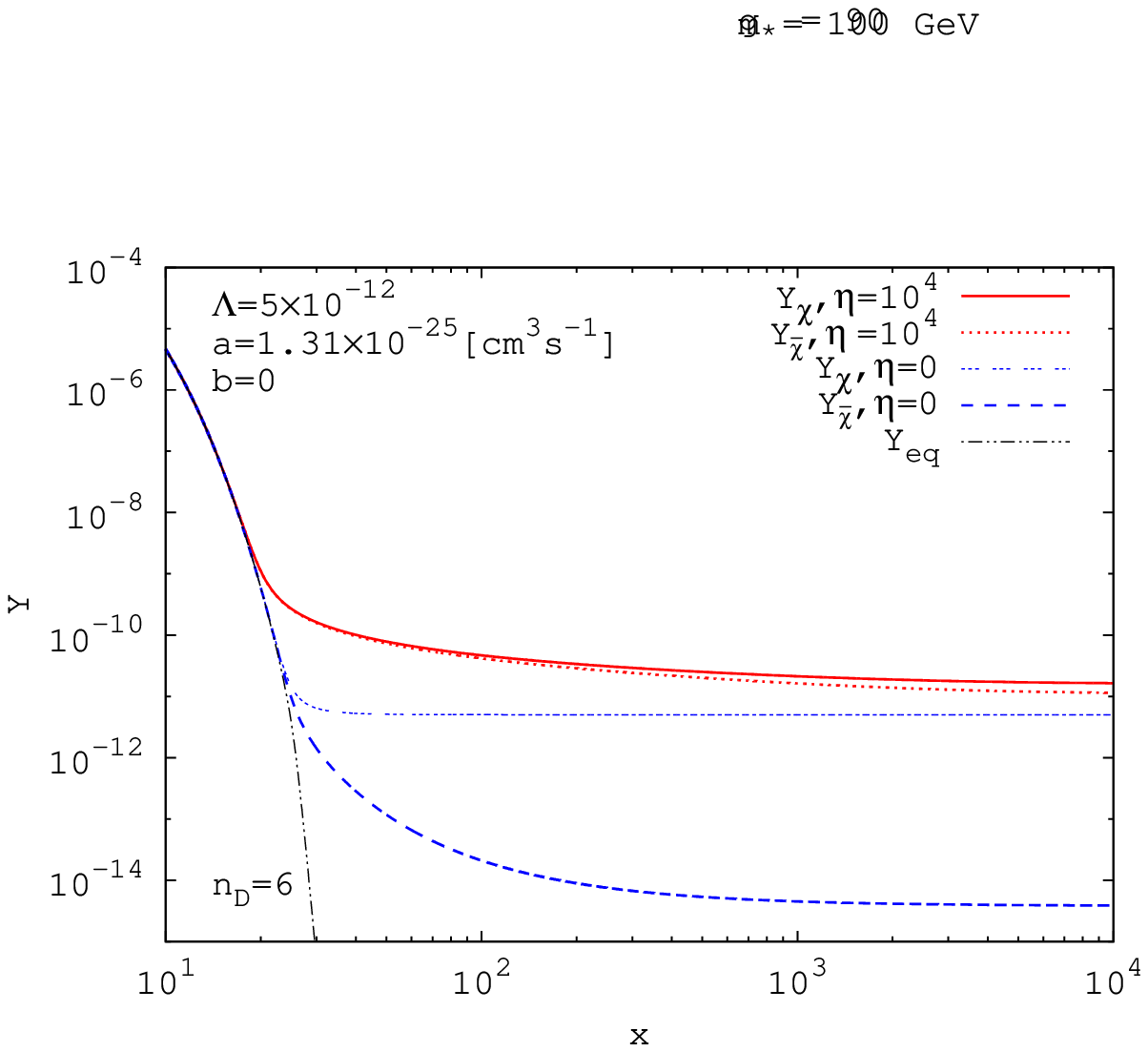}
    \put(-115,-12){(b)}
     \vspace{0.5cm}
    \hspace*{-0.5cm} \includegraphics*[width=8cm]{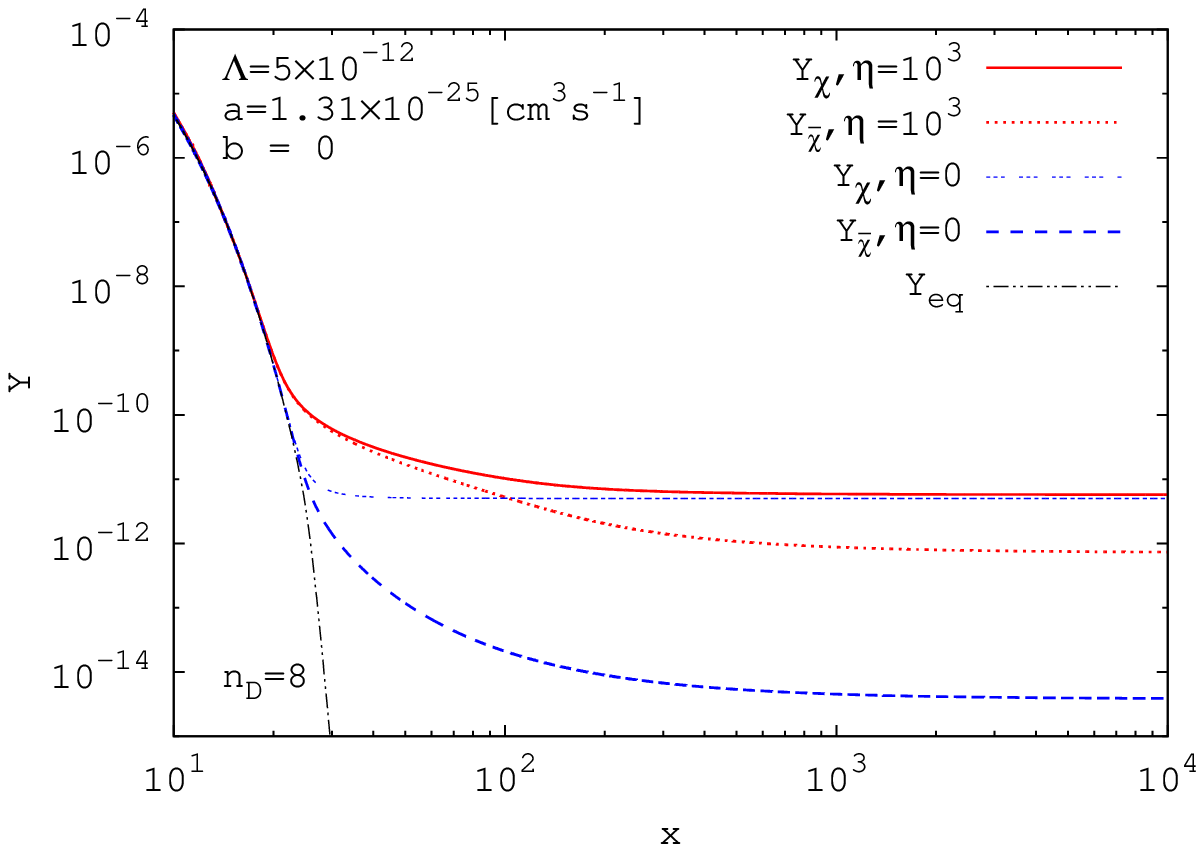}
    \put(-115,-12){(c)}
    \hspace*{-0.5cm} \includegraphics*[width=8cm]{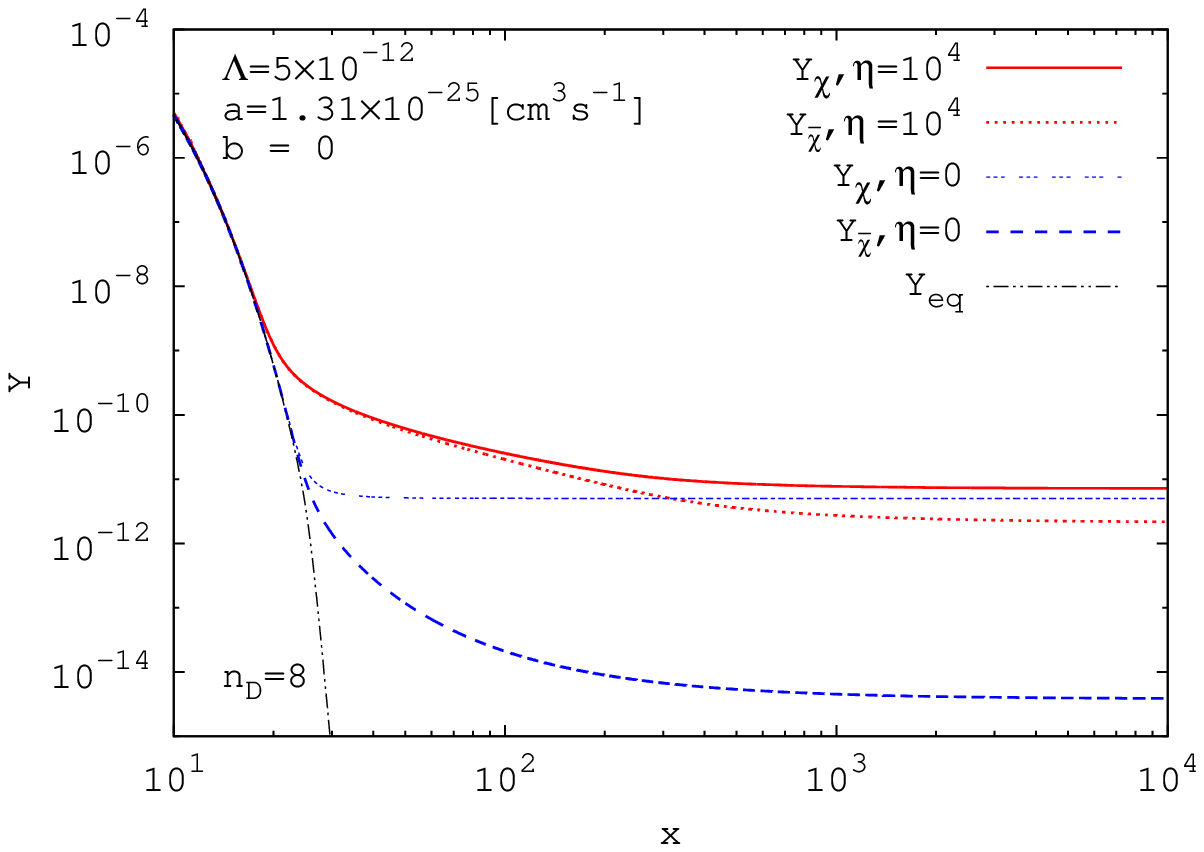}
    \put(-115,-12){(d)}
    \caption{\label{fig:a} \footnotesize Evolution of $Y_{\chi}$ and
     $Y_{\bar\chi}$ as a function of the inverse--scaled temperature $x$ for
     $n_D = 6$ and $n_D = 8$. $\eta = 10^3$ for panels (a), (c) and 
     $\eta = 10^4$ for panels (b), (d).  Here
     $ \Lambda = 5 \times 10^{-12} $, $m = 100$ GeV, $g_{\chi} = 2$,
     $g_* = 90$, $ a = 1.31 \times 10^{-25} $ ${\rm cm}^3 \,{\rm s}^{-1}$, 
     $b = 0$, $x_0=25$.}
  \end{center}
\end{figure}

Fig.\ref{fig:a} demonstrates the evolution of 
the abundances for particle and anti--particle for different $n_D$ and
$\eta$. These plots are based on the numerical solutions of 
Eqs.(\ref{eq:Yc}), (\ref{eq:Ycbar}). As we mentioned earlier 
$n_D = 6$ corresponds to the quintessence scenario with kination phase while 
$n_D = 8$ is for the brane world cosmology or the late inflaton decay model. We
find the abundances for particle and anti--particle are increased in those 
models. The increase depends on the enhancement factor $\eta$, when $n_D = 6$, 
we see that the increase is larger for $\eta =10^4$ in panel 
(b) compared to $\eta =10^3$ in panel (a). The situation is the same for 
$n_D = 8$ in panels (c) and (d).
On the other hand, it is shown the increase of the relic abundance is larger 
for smaller $n_D$ when $\eta$ takes the same value. 
The reason for the increases of the relic abundances for particle 
and anti--particle are the enhanced Hubble expansion rate in those modified 
cosmological models. The enhanced Hubble expansion rate of the universe leads 
to the early decay of the asymmetric Dark Matter particles from the 
equilibrium state. Therefore, there are more particles and anti--particles 
in the modified scenarios with respect to the standard cosmological scenario. 

\subsection{Analytical solutions}

First, we find the analytic solution for Eq.(\ref{eq:Ycbar}), and then 
obtain the solution for Eq.(\ref{eq:Yc}) using the relation 
$ Y_{\chi} - Y_{\bar\chi} = \Lambda $. In terms of 
$\Delta_{\bar\chi} = Y_{\bar\chi} - Y_{\bar\chi,{\rm eq}}$,
Eq.(\ref{eq:Ycbar}) is written as
\begin{equation} \label{eq:delta}
\frac{d \Delta_{\bar\chi}}{dx} = - \frac{d Y_{\bar\chi,{\rm eq}}}{dx} -
\frac{\kappa \langle \sigma v \rangle}{x^2}~
     \left[ 1 + \eta \left( \frac{x_0}{x} \right)^{n_D -4} \right]^{-\frac{1}{2}}\,
\left[\Delta_{\bar\chi}(\Delta_{\bar\chi} + 2 Y_{\bar\chi,{\rm eq}})
      + \Lambda \Delta_{\bar\chi}   \right]\, .
\end{equation}
Repeating the same method which is mentioned in \cite{Iminniyaz:2011yp},
we obtain
\begin{equation} \label{bardelta_solu}
      \Delta_{\bar\chi} \simeq \frac{2 x^2 P 
   \left[ 1 + \eta \left( x_0/x \right)^{n_D - 4} \right]^{\frac{1}{2}}}   
      {\kappa \langle \sigma v \rangle\,({\Lambda}^2 + 4 P)}\,,
 \end{equation}
for high temperature. The inverse--scaled freeze--out temperatures 
$x_F$ and $\bar{x}_F$ are determined by this solution \cite{Iminniyaz:2011yp}.
For sufficiently low temperature, we obtain 
\begin{equation} \label{eq:barY_cross}
Y_{\bar\chi}(x \rightarrow \infty) =  \Lambda\,\left\{ \,
  {\rm exp} \left[\, 1.32\, \Lambda \, m M_{\rm Pl}\,
 \sqrt{g_*} \, I(\bar{x}_F)\,   \right] -1\, \right\}^{-1} \,,
\end{equation}
where
\begin{eqnarray}
    I(\bar{x}_F) & =& \int^{\infty}_{\bar{x}_F} \frac{ \langle \sigma v \rangle }
              {x^2}\left[ 1 + \eta 
   \left( \frac{x_0}{x} \right)^{n_D -4} \right]^{-\frac{1}{2}}\, dx.
\end{eqnarray}
The relic abundance for $\chi$ particle is 
\begin{equation}\label{eq:Y_cross}
      Y_{\chi}(x \rightarrow \infty) = \Lambda\,
 \left\{\, 1 - \exp \left[\, - 1.32\,  \Lambda\, m M_{\rm Pl}\,
 \sqrt{g_*} \,  I(x_F) \,   \right]\, \right\}^{-1}\,.
\end{equation}
For $x_F = \bar x_F$, the solutions for $Y_{\chi}$ and $Y_{\bar\chi}$ 
match with the constraint 
$Y_{\chi} - Y_{\bar\chi} = \Lambda$. The final abundance is expressed as
\begin{equation}
\Omega_\chi h^2 = \frac{\rho_{\chi} }{\rho_{\rm  crit}} h^2
               =\frac{m s_0 Y_{\chi}(x \to \infty) h^2}{\rho_{\rm  crit}}\,,
\end{equation}
where 
\begin{equation} \label{s_0}
   \rho_{\chi}=n_{\chi} m\,, \,\,\,\,
    \rho_{\rm crit} = 3 M_{\rm Pl}^2 H_0^2\,,\,\,\,\, 
     s_0 = 2.9 \times 10^3~{\rm cm}^{-3}\,.
\end{equation}
The predicted present total Dark Matter relic density is then
\begin{eqnarray} \label{omega}
 \Omega_{\rm DM}  h^2 & = & 2.76 \times 10^8~ m \left[ Y_{\chi}~(x
  \rightarrow \infty) + Y_{\bar\chi}~(x \rightarrow \infty) \right].
\end{eqnarray}
According to the assumption that the deviation
$\Delta_{\bar\chi} $ is of the same order of the equilibrium value of
$Y_{\bar\chi}$, the freeze--out temperature for $\bar{\chi}$ is fixed using
the equality $\xi Y_{\bar\chi,{\rm eq}}( \bar{x}_F) = \Delta_{\bar\chi}( \bar{x}_F) $
where $\xi$ being a constant, usually $\xi = \sqrt{2} -1$ \cite{standard-cos}.

\section{Constraints}
The Dark Matter relic density is given by Planck date as in
Eq.(\ref{eq:pdata}).
\begin{figure}[h]
  \begin{center}
    \hspace*{-0.5cm} \includegraphics*[width=8.7cm]{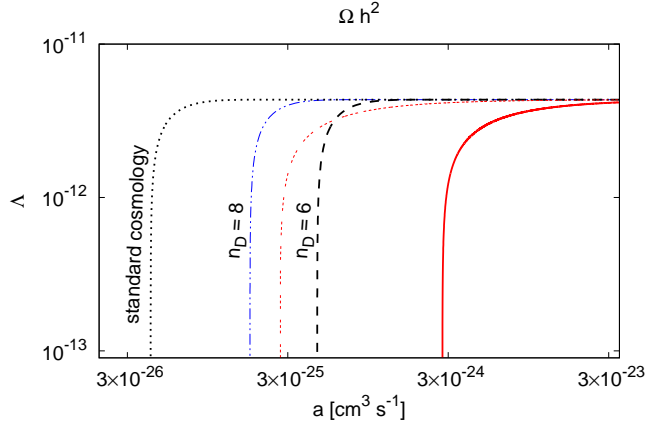}
     \caption{\label{fig:d} \footnotesize
    The allowed region in the $(a,\Lambda)$ plane for different $n_D$ and 
    standard cosmology when the Dark Matter relic density 
    $\Omega h^2 = 0.1199$. Here 
    $m = 100$ GeV, $g_{\chi} = 2$ and $g_* = 90$, $x_0=25$, $\eta = 10^3$.
    The three dot--dashed (red) line is
    the limiting cross section which 
    corresponds to the Fermi--LAT bounds for 
    $\chi\bar\chi \rightarrow b\bar{b}$ channel 
    and the thick (red) line is for 
    $\chi\bar\chi \rightarrow \mu^{\dagger} \mu^{-}$ channel 
    \cite{Ackermann:2013yva}.  }
    \end{center}
\end{figure}
The contour plot of the $s$--wave 
annihilation cross section $a$ and asymmetry factor $\Lambda$ is 
displayed in 
Fig.\ref{fig:d} when the Dark Matter relic 
density is $\Omega_{\rm DM} h^2 = 0.1199$. Here the 
dashed (black) line is for $n_D = 6$, the double dot--dashed (blue) line is for 
$n_D =8$ and the dotted (black) line is for the standard cosmology. 
We found the cross sections are larger for the modified cosmological models
 with respect to the standard cosmology. The relic density is larger in 
modified cosmology due to the enhanced
Hubble rate, therefore the cross section should be increased in 
order to satisfy the observed range of the Dark Matter relic 
density. Of course the increased value of the cross section depends on the 
parameter $n_D$. For smaller $n_D$ (dashed line), the larger cross 
section is needed to satisfy the observed range of Dark Matter relic density, 
for example, the cross section changes from 
$ 4.6 \times 10^{-25}$ ${\rm cm}^3\,{\rm s}^{-1}$ to 
$ 3.5 \times 10^{-23}$ ${\rm cm}^3 \,{\rm s}^{-1}$ for $n_D = 6$ and 
$ 1.8 \times 10^{-25}$ ${\rm cm}^3 \,{\rm s}^{-1}$ to  
$ 3.5 \times 10^{-23}$ ${\rm cm}^3 \,{\rm s}^{-1}$ for $n_D = 8$. Here 
$\Lambda$ takes the value from  
$ \Lambda = 9.0 \times 10^{-14}$ to $\Lambda = 4.4 \times 10^{-12}$. We can 
analyze this
from panels (a) and (c) in Fig.\ref{fig:a}. For the same $\eta$,
the increase of relic abundance is larger for $n_D=6$ with respect to the case
$n_D=8$, therefore the annihilation cross section should be 
large to satisfy the observed range of the Dark Matter abundance for 
smaller $n_D$. 

Since the annihilation cross sections of the asymmetric Dark Matter
particles in the modified cosmological scenarios are increased, then the
annihilation rate is enhanced and the indirect detection is possible for
the asymmetric Dark Matter particles which is originally assumed to be
detected only through the direct detection due to the suppressed abundance of
the anti--particles. Therefore we can use the 
Fermi Large Area Telescope (Fermi--LAT) \cite{Ackermann:2013yva} data to find 
the limiting cross sections for the asymmetric Dark Matter in nonstandard 
cosmological scenarios. In the case of the symmetric Dark Matter, the 
annihilation rate is \cite{Gelmini:2013awa}
%
\begin{equation}\label{gamma:s}
      \Gamma_{\rm sym} = \frac{1}{2}\langle \sigma_{\rm self} v \rangle 
      \frac{\rho^2_{\rm DM}}{m^2}\,,
\end{equation} 
where $\rho_{\rm DM} = \rho_{\rm crit} \Omega_{\rm DM} $ and 
$\langle \sigma_{\rm self} v \rangle$ is the thermal average of the self
annihilation cross section. For asymmetric Dark
Matter, the annihilation rate is
\begin{equation}\label{gamma:as}
      \Gamma_{\rm asym} = \langle \sigma v \rangle 
       \frac{\rho_{\chi}\rho_{\bar\chi}}{m^2}\,,
\end{equation}
here we use $\rho_{\chi} + \rho_{\bar\chi} = \rho_{\rm DM}$ and 
$n_{\chi}=\rho_{\chi}/m$, then Eq.(\ref{gamma:as}) becomes
\begin{equation}\label{gamma:f}
      \Gamma_{\rm asym} = \langle \sigma v \rangle \frac{\rho^2_{DM}}{m^2}
                      \frac{Y_{\chi}Y_{\bar\chi}}{(Y_{\chi}+Y_{\bar\chi})^2}\,.
\end{equation}
According to the analysis of \cite{Gelmini:2013awa}, the ratio of 
$\Gamma_{\rm asym}$ and $\Gamma_{\rm Fermi}$ must be less than one,
\begin{equation}
      \frac{\Gamma_{\rm asym}}{\Gamma_{\rm Fermi}} = 
      \frac{\langle \sigma v \rangle }{\langle \sigma v \rangle_{\rm Fermi}}
      \frac{2Y_{\chi}Y_{\bar\chi}}{(Y_{\chi} + Y_{\bar\chi} )^2} < 1\,,
\end{equation}
where $\langle \sigma v \rangle_{\rm Fermi}$ is the derived cross section
bound from the Fermi--LAT data.
Then using Eq.(\ref{omega}), we obtain
\begin{equation}\label{eq:bound}
      \frac{\Omega_{\rm DM}h^2}{2.76 \times 10^8 m}
      \left(1 -2\frac{\langle \sigma v\rangle_{\rm Fermi}}
      {\langle \sigma v\rangle} \right)^{1/2} < \Lambda\,.
\end{equation} 
Fermi--LAT provied the following uppor bounds on the annihilation
cross section $ a = 1.34 \times 10^{-25}$ ${\rm cm}^3 \,{\rm s}^{-1}$ for 
$\chi\bar\chi \rightarrow b\bar{b}$ channel and 
$a = 1.38 \times 10^{-24}$ ${\rm cm}^3 \,{\rm s}^{-1}$ for 
$\chi\bar\chi \rightarrow \mu^{\dagger} \mu^{-}$ channel while 
$m = 100$ GeV \cite{Ackermann:2013yva}. Applying Planck data
Eq.(\ref{eq:pdata}) and Fermi-LAT data to Eq.(\ref{eq:bound}) together,
we obtain the limiting bound for the annihilation cross sections in the 
modified cosmological scenarios. We find
when $ \Lambda = 9 \times 10^{-14}$, for $\chi\bar\chi \rightarrow b\bar{b}$
channel, the corresponding constrained cross sections are
$\langle\sigma v \rangle \leq 2.7\times 10^{-25} $ 
${\rm cm}^3 \,{\rm s}^{-1}$(three dot--dashed line in Fig.\ref{fig:d}); 
for $\chi\bar\chi \rightarrow \mu^{\dagger}\mu^{-}$ channel, it is 
$\langle\sigma v\rangle \leq 2.8 \times 10^{-24}$ 
${\rm cm}^3 \,{\rm s}^{-1}$ (thick red line in Fig.\ref{fig:d}).
\begin{figure}[h]
  \begin{center}
    \hspace*{-0.5cm} \includegraphics*[width=8.7cm]{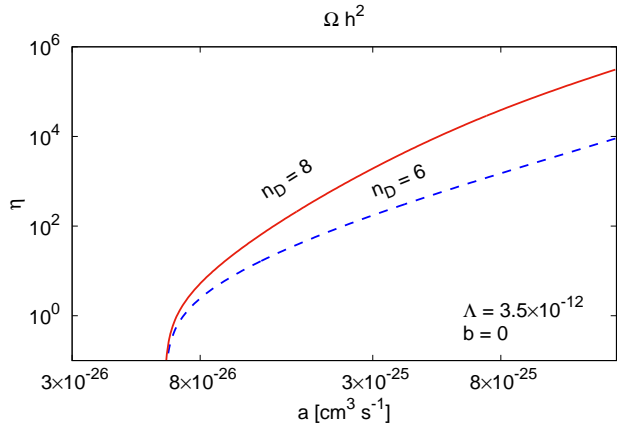}
     \caption{\label{fig:e} \footnotesize
    The allowed region in the $(a,\eta)$ plane for $b=0$ when the Dark Matter 
    relic density $\Omega h^2 = 0.1199$. Here we
    take $m = 100$ GeV, $ \Lambda = 3.5 \times 10^{-12}$, $g_{\chi} = 2$ and 
    $g_* = 90$, $x_0=25$. The dashed (blue) line is for $n_D=6$ and the thick 
    (red) line is for $n_D=8$.  }
    \end{center}
\end{figure}
Fig.\ref{fig:e} demonstrates the relation between the $s$--wave annihilation 
cross section $a$ and the enhancement factor $\eta$ when $\Omega h^2 = 0.1199$. 
In this plot, we take the value for asymmetry factor as
  $\Lambda=3.5 \times 10^{-12}$ which is in the range of the values to obtain
  the  observed Dark Matter abundance. It is indeed shown in
  Fig.\ref{fig:d}. Here the thick (red) line is for $n_D = 8$ and the 
dashed (blue) line is for $n_D =6$. While $n_D = 8$, for the cross sections 
from $6.2 \times 10^{-26}$ ${\rm cm}^3\, {\rm s}^{-1}$ to
$1.9 \times 10^{-24}$ ${\rm cm}^3 \,{\rm s}^{-1}$, $\eta$ changes from $0.1$ to 
$3.1 \times 10^5$ and to $9.0 \times 10^3$ when $n_D = 6$. For the same cross 
section, the enhancement factor $\eta$
is increased for larger $n_D$. We can understand the reason
from Fig.\ref{fig:a}. In the frame (b) and (c) of Fig.\ref{fig:a}, the cross 
section is same for $n_D = 6$ and $n_D =8$. We found for 
smaller $n_D$, there are more increase of asymmetric Dark Matter relic density 
for $\eta = 10^4$ in panel (b) than the case for $\eta = 10^3$ in panel (c). 
In order to fall the observed range of Dark Matter relic density, $\eta$ 
should be decreased (increased) for small (large) value of $n_D$. 

In this work, we use the observed Dark Matter abundance to derive
  the constraints on the enhancement factor for the fixed asymmetry factor 
$\Lambda$. In the quintessence model ($n_D=6$), the constraint
on the enhancement factor from BBN is $\eta<30$ \cite{Dutta:2010cu} 
 which matches partly with the limit $\eta< 9.1\times 10^3$ that we obtained. 
In paper \cite{Dutta:2010cu}, the
 bound on $\eta$ was obtained at the temperature $T= 10$ MeV, here we take this
 bound as an example of a reference. As we mentioned
  earlier $n_D=8$ corresponds to the brane Randall-Sundrum II model 
\cite{Randall:1999vf}, which in our notations represents  
$\eta = \rho_{\rm rad}(T_0)/(2\lambda)$ with $\lambda$ being the 
tension of the brane which is related to the
5--dimensional Planck mass $M_5$ by $\lambda = 48 \pi M_5^6/ M^2_{\rm P}$, here
$M_P$ is the 4--dimensional Planck mass \cite{Meehan:2014zsa}. 
From Fig.\ref{fig:e}, we see that $\eta < 3.1\times 10^5$, this implies 
$\lambda > 1.9 $ GeV$^4$ and it leads $M_5 > 10^3$ TeV. The BBN bound is 
$\lambda >1$ MeV$^4$ and so $ M_5 > 30$ TeV \cite{Durrer:2005dj}.

\section{Summary and conclusions}

We studied the relic density of asymmetric Dark Matter particles 
in the modified cosmological scenarios. We have no observational evidence for 
the early pre--BBN universe. There are different cosmological 
models which predicted the Hubble expansion rate before BBN maybe faster or 
slower than the standard expansion law. If the Hubble expansion rate is 
changed in the modified 
cosmological scenarios, it affects the relic density of asymmetric Dark Matter 
particles. We discussed in which extent the modification of the
Hubble expansion rate affects the relic density of asymmetric Dark Matter in 
the modified cosmological models. 

The Hubble expansion rate of the universe is increased in the modified 
cosmological scenarios like quintessence with kination phase, brane world 
cosmology or the late inflaton decay model and etc. For generality we found the 
parameterization of the enhancement 
of the Hubble expansion rate. Then we calculated the relic
density of asymmetric Dark Matter particles in the modified cosmological 
scenarios where the expansion rate is faster than the
standard cosmological scenarios. We found the resulting relic 
abundances are increased. Since the expansion rate of the 
universe is enhanced during the era of asymmetric Dark Matter decoupling, the 
asymmetric Dark Matter particles decouple from the thermal equilibrium earlier
than the standard cosmological scenario and there are 
increased relic densities for particles and anti--particles. The increases 
depend on the enhancement factor $\eta$ as well as the cross section and the 
factor $n_D$. While using Planck date for the relic density of asymmetric Dark
Matter, we found the cross sections are boosted for the
modified cosmological models in order to
satisfy the observed Dark Matter relic density. From the contour plot of
($a, \eta$), we found $\eta$ is decreased (increased) for 
small (large) value of $n_D$ when the cross section is same for different 
$n_D$.

The generalised formula obtained provides the relic density calculation of the
asymmetric Dark Matter particles for the different modified cosmological 
scenarios where the expansion rate is enhanced. In the standard cosmological 
model, the abundance of minority component of asymmetric Dark Matter 
is suppressed. For the modified cosmological models, we obtained 
the enhanced relic densities both for particles and anti--particles, of course 
the increased amount of the relic abundance depends on the size of the 
annihilation cross section. 
The annihilation cross section should be boosted in order to provide the
observed range of Dark Matter relic density (\ref{eq:pdata}). The corresponding 
annihilation rate is increased too. This effect is contrary to the general 
belief that the asymmetric Dark Matter is only detected by direct detection. In
modified cosmological scenarios, we 
have the possibility to detect the asymmetric
Dark Matter particle in indirect way. The asymmetric Dark Matter particle 
properties may be constrained by the indirect detection signals
\cite{Ackermann:2013yva}. 
In addition to the observational Dark Matter relic density 
\cite{Ade:2015xua}, there are collider, direct and indirect detection signals, 
using those data the parameter space is examined before BBN.

\section*{Acknowledgments}

The work is supported by the National Natural Science Foundation of China
(11365022) and (11765021).

\end{document}